\documentclass[prx,aps,showpacs,twocolumn,preprintnumbers,amsmath,amssymb,longbibliography]{revtex4-2}
\usepackage[english]{babel}
\usepackage[utf8]{inputenc}
\usepackage[T1]{fontenc}
\usepackage{amsmath,amssymb,bbm,mathrsfs,mathtools,multirow,diagbox,xcolor,%
  graphicx,tabularx,comment,amsfonts,dsfont,lettrine,units,comment}
\usepackage{graphicx}
\usepackage[colorlinks=true,allcolors=blue]{hyperref}
\usepackage{etoolbox}
\AtBeginEnvironment{thebibliography}{\hypersetup{colorlinks=false}}
\usepackage{bookmark}
\usepackage{xcolor}
\usepackage{chngcntr}
\usepackage{braket}
\usepackage{nicefrac}
\usepackage{bm}
\usepackage{bbm}
\usepackage{braket}
\usepackage{lineno}
\usepackage{amssymb}
\usepackage{array}
\newcolumntype{C}{>{$}c<{$}}
\usepackage{newtxtext} 
\usepackage{ulem}
\usepackage{graphicx}

\def\ket#1{|#1\rangle }
\def\bra#1{\langle #1 |}

\DeclareMathOperator{\Tr}{Tr\,}
\DeclareMathOperator{\sech}{sech}

\usepackage{xstring} 

\begin{document}
\title{Diamagnetic Meissner response of odd-frequency superconducting pairing from quantum geometry}

\author{Ankita Bhattacharya}
\email{ankita.bhattacharya@physics.uu.se}
 \affiliation{Department of Physics and Astronomy, Uppsala University, Box 516, 
751 20 Uppsala, Sweden}
\author{Annica M. Black-Schaffer}
 \affiliation{Department of Physics and Astronomy, Uppsala University, Box 516, 
751 20 Uppsala, Sweden}

\begin{abstract}
We investigate the role of quantum geometry in the Meissner response for odd-frequency superconducting pairs in multiband systems. Odd-frequency pairing is traditionally associated with a paramagnetic Meissner response, which raises questions about the stability of the superconducting phase, especially in multiband systems where odd-frequency pairing is ubiquitous. 
Using analytical calculations in a general two-band, we show that the quantum geometric contribution to the Meissner response from odd-frequency pairs is always diamagnetic for its interband processes,  while intraband processes always yield a paramagnetic response. 
With odd-frequency pairing itself generated by interband pairing, an overall diamagnetic response may often be anticipated. 
We confirm these results with numerical calculations of models with both flat and dispersive bands. 
In flat band systems, where geometric effects dominate, the diamagnetic odd-frequency response can even exceed the even-frequency contribution, making odd-frequency pairs the primary source of the diamagnetic Meissner response. In a dispersive two-band system with finite quantum geometry, we similarly find a robust diamagnetic contribution from odd-frequency pairing, even when the total response turns paramagnetic due to even-frequency contributions. These results establish that quantum geometry stabilizes odd-frequency superconductivity and also identify flat-band materials as candidates for realizing odd-frequency superconductivity with a diamagnetic Meissner effect.

\end{abstract}
\date{\today}
\maketitle
\section{Introduction} 
\label{sec:introduction}
The quantum geometry of electrons has recently attracted widespread attention~\cite{Rossi21,Torma23}, especially for its decisive role in flat band superconductors~\cite{PT15,TPB22,PHT25}, in particular in twisted bilayer graphene~\cite{TBG_SC18,Rossi19,Bernevig20,TPB22,Park22,Tanaka25}. 
The impact of quantum geometry is also not limited to flat band superconductors. The quantum geometric tensor (QGT), or Fubini-Study metric, determines the distance between two quantum states and depends on the geometry of the eigenstate space~\cite{Rossi21,Torma23}. The imaginary part of the QGT, well-known as the Berry curvature, measures the phase distance between states and not only determines the topology of the system, but also enters in various linear and nonlinear electronic transport phenomena~\cite{AHE10,Qian10,Sodeman15}. The real part of the QGT, known as the quantum metric, is crucial in various higher-order transport processes~\cite{Gao14,THE21,BB25}, resonant optical responses~\cite{Gao19,Ahn22}, and fractional Chern insulators~\cite{Roy14,Bergholtz22}, along with generating a finite superfluid weight in flat band superconductors~\cite{PT15,PHT25,TBG_SC18,TPB22}.

Since the quantum geometry arises due to the structure of the Bloch functions in reciprocal space, it becomes pertinent only for multiband systems, where the bands consist of contributions from different orbitals, valleys, layers, or similar degrees of freedom (DoFs). In particular, the quantum metric enters in various physical processes through the velocity operator, specifically through the off-diagonal, or interband, velocity~\cite{Rossi21,Torma23,Chen23,Huang25}, which depicts a coherent motion of electrons between different Bloch bands.  For superconductors, this off-diagonal velocity gives an additional contribution to the superfluid weight, beyond the traditional contribution from the intraband velocity, or equivalently, electron dispersion. This is particularly crucial in flat band superconductors, since their vanishing dispersion leads to a zero conventional superfluid weight ~\cite{PT15,PHT25,TBG_SC18,TPB22}.
 
Superconductors possessing multiple DoFs also ubiquitously host odd-frequency superconductivity~\cite{BB13,LB19,TCB20,CB22}. Odd-frequency superconductivity is an extraordinary superconducting pairing state where the two electrons forming a Cooper pair, are paired at different times with an odd relative time difference or, equivalently, an odd-frequency dependence~\cite{BVE05,LB19}. The explicit presence of time or frequency as a DoF directly influences the symmetry properties of the superconducting state, e.g., an odd-frequency spin-triplet state can still have $s$-wave spatial symmetry \cite{Efetov03}, since the total Cooper pair wave function is odd under the exchange of all quantum numbers of the constituent electrons, also including time/frequency~\cite{BVE05,LB19}. Odd-frequency pairing has therefore been extensively studied in proximity-induced superconductivity in superconducting-magnet heterostructures~\cite{Efetov03,Efetov05,Golubov13,PB14, PRX15,CTB20, Kim21}, where odd-frequency spin-triplet $s$-wave superconducting pairs are generated.

Beyond heterostructures, odd-frequency superconductivity has also been predicted in the bulk of multiband systems \cite{BB13,KB15,TCB20,TR16,CB21,PB19,Ebisu16,PB17,Triola16,CTB21}. Here, the band index provides an additional DoF, such that e.g.~spin-singlet $s$-wave pairs with odd-frequency and odd-interband symmetries are allowed.
In fact, the condition for obtaining odd-frequency bulk multiband superconductivity is simply that interband pairing exists, along with the participating bands having different dispersions. Such interband pairing is easy to achieve in any multiorbital materials, as it only requires a finite interorbital hybridization, along with an asymmetry in the intraorbital pairing, i.e., the intraorbital pairing amplitudes are different in the participating orbitals~\cite{BB13,PB21}. 
Consequently, many multiband superconductors have been suggested to host odd-frequency pairing~\cite{KB17,TB18,PB21,MT21,DPB21,BBTI13,Schmidt20}. 

One of the defining features of superconductivity is the diamagnetic Meissner response, i.e., expulsion of magnetic fields from the superconductor, which is intimately tied to a positive superfluid weight~\cite{MO93}.
However, odd-frequency superconductors have instead been shown to host a paramagnetic Meissner response due to a negative superfluid weight~\cite{Hoshino14,AS15,PRX15,LSR17,KPG20}, meaning the superconductor instead attracts magnetic field, thus making it unstable in the presence of an external magnetic field. A paramagnetic Meissner response is not a problem for odd-frequency proximity-induced superconductivity in heterostructures, since they are built up of a conventional bulk superconductor, which is stable in magnetic fields. However, a paramagnetic Meissner response from odd-frequency pairing in bulk superconductors raises the question about the ultimate stability of such superconductors. 

Only a few works have so far attempted to investigate the odd-frequency contributions to the Meissner response in multiband systems. 
In simple multiband systems, a paramagnetic Meissner effect has been found from odd-frequency pairing, thereby reducing stability and the superconducting transition temperature \cite{AS15}. On the other hand, for the proposed unconventional superconducting state in the doped topological insulator $\mathrm{Bi}_2\mathrm{Se}_3$~\cite{Schmidt20}, where both strong spin-orbit coupling and a linear band dispersion are present, odd-frequency pairing has been shown to instead give rise to a diamagnetic Meissner response. Also, more  generic systems hosting an inverted band structure with quadratically dispersive bands have been found to possess a diamagnetic Meissner response from its odd-frequency pairs~\cite{PB21}. Thus, the Meissner effect in multiband superconductors seems to be highly dependent on the band dispersion. 
However, no work, to the best of our knowledge, has investigated the role of quantum geometry for the odd-frequency contributions to the Meissner response. This question is crucially important for flat band superconductors, where a finite quantum geometry is the only way to generate a positive superfluid weight and a diamagnetic Meissner response \cite{PT15, Rossi19, Rossi21, Torma23, Thumin23,TPB22,Barlas24,Hosur25,Huang25}.

This leads us to the the main question we intend to answer in this work: 
Is the geometric contribution to the Meissner effect dia- or paramagnetic for odd-frequency pairing? 
Since the geometric contribution is dominating in flat-band systems, this also directly leads to two additional questions: 
Does odd-frequency pairing in flat-band systems give a dia- or paramagnetic Meissner effect and can the odd-frequency response there even dominate the even-frequency response?

In this work, we address these questions by studying the role of quantum geometry for both the even- and odd-frequency contributions of the Meissner response for general two-band systems. We show analytically that the geometric contribution from odd-frequency pairing is always diamagnetic for the interband processes involved in the Meissner response, but always paramagnetic for the intraband processes. Since interband pairing is the origin of odd-frequency pairing, we may expect the interband contribution to often dominate over the intraband contribution, such that the overall geometric odd-frequency Meissner response is diamagnetic. 
We numerically confirm these results by studying a prototype model hosting a flat band and a dispersive band. We even find that the odd-frequency contribution to the Meissner effect can easily be larger than the even-frequency contribution, thus making odd-frequency pairs the primary driver for a diamagnetic Meissner effect in flat-band systems. 
To complement the flat band results, we also study a model hosting two dispersive bands with a finite quantum geometry. Here we also find a diamagnetic Meissner response from the odd-frequency pairing, primarily of geometric origin. Even when the system has an overall paramagnetic Meissner response, the odd-frequency pairing is still contributing diamagnetically, and it is instead the even-frequency pairs driving the paramagnetic response. To summarize, the geometric contribution to the Meissner response for odd-frequency pairing is expected to be diamagnetic, generated by dominating interband processes. In flat band systems the odd-frequency diamagnetic Meissner response can even dominate the even-frequency contribution. This conclusively answers the questions raised above. 
Our results also point to flat band materials being a promising route for purely odd-frequency superconductivity with a stabilizing diamagnetic Meissner response. 

The remainder of the work is organized as follows. In Sec.~\ref{sec:theory_model}, we first introduce two multiorbital models to describe the normal state of prototype flat-band and dispersive systems, respectively. We then discuss the superconducting pairing correlations and find the condition for odd-frequency pairing in Sec.~\ref{sec:theory_SC}. General theory of the Meissner effect is presented in Sec.~\ref{sec:theory_Meissner}. We report our analytical findings generic to all multiband superconductors in Sec.~\ref{results: GA} and complement with numerical results for the flat-band model in Sec.~\ref{results:Model A} and for the dispersive model in Sec.~\ref{results:Model B}. We conclude our work in Sec.~\ref{sec:conclusions}.

\section{Theoretical framework} 
\label{sec:theory}
In order to investigate the role of quantum geometry in the Meissner response for both even- and odd-frequency pairing, we consider a generic two band systems. For the numerical studies we choose two simple two-band systems, \textit{Model A} featuring one flat band and one dispersive band, and \textit{Model B} featuring two dispersive bands. Overall, to obtain sizable bulk odd-frequency superconducting pairing, only two ingredients are sufficient, namely, the presence of multiple orbitals with finite inter-orbital hybridization and the presence of a pairing asymmetry, i.e., the strength of the superconducting pairing should be different in different orbitals \cite{BB13,TCB20,PB21}. Notably, these are easy conditions to satisfy in multiorbital systems with pairing assumed to be orbital-based. Then, in the band basis, where the kinetic Hamiltonian is diagonal, the condition to obtain the bulk odd-frequency pairing translates to simply having finite interband pairing \cite{BB13,TCB20,PB21}, as we also demonstrate below. Below, we first introduce the two models we used in our numerical studies, then focus on superconductivity and the Meissner effect.

\subsection{Normal state}
\label{sec:theory_model}
For the numerical studies we use two different models, Model A and Model B, presented below.
\subsubsection{Model A}
\begin{figure}[thb]
    \centering
    \includegraphics[width=0.91\columnwidth]{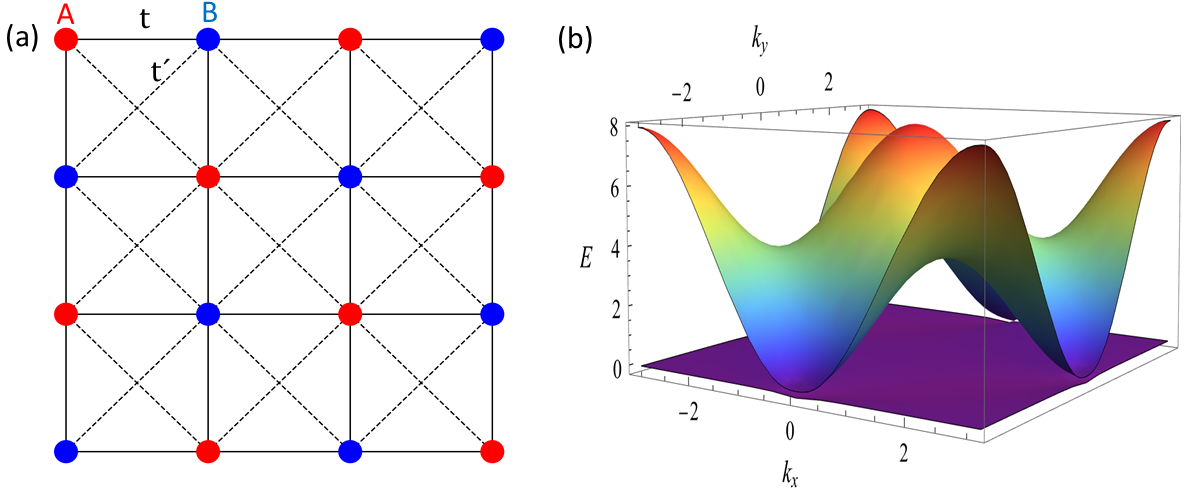}
    \caption{ Model A described by the Hamiltonian Eq.~(\ref{eq.H1}). (a) Schematic diagram of the tight-binding square lattice containing two non-equivalent lattice sites $A$ (red) and $B$ (blue) and hopping parameters $t$ and $t^\prime$. (b) Energy dispersion for $t=t^\prime=1$, $\delta=0.01$, and $\mu=-1.95$.}
    \label{fig:BS_CB}
\end{figure}
For model A we consider the likely simplest two-orbital lattice model in $2$D that hosts a flat band: a tight-binding Hamiltonian on a checkerboard lattice containing two non-equivalent sites per unit cell such that the kinetic Hamiltonian in the basis $ (c_{A, \mathbf{k}}, c_{B, \mathbf{k}})^T$ reads 
\begin{align}
  H^\text{A}_{\mathbf{k}} = 
   \begin{pmatrix}
   \epsilon_{A, \mathbf{k}} & \epsilon_{AB, \mathbf{k}}\\
   \epsilon_{BA, \mathbf{k}} &   \epsilon_{B, \mathbf{k}} 
      \end{pmatrix} ,
      \label{eq.H1}
\end{align}
where $ \epsilon_{A, \mathbf{k}}= -2 t^\prime (\cos{k_x} \cos{k_y} + \sin{k_x} \sin{k_y}) + \delta $, $ \epsilon_{B, \mathbf{k}}= -2 t^\prime (\cos{k_x} \cos{k_y} - \sin{k_x} \sin{k_y}) - \delta $, and $ \epsilon_{AB, \mathbf{k}}= \epsilon_{BA, \mathbf{k}}= 2 t (\cos{k_x} + \cos{k_y})$. Here, $ \epsilon_{A/B, \mathbf{k}}$ represents intra-sublattice hopping, where $t^\prime$ connects next-nearest neighbors of the same sublattice, while the off-diagonal $\epsilon_{AB, \mathbf{k}}$ term is the inter-sublattice hopping with $t$ connecting nearest neighbor sites of different sublattices, see Fig.\,\ref{fig:BS_CB}(a). The parameter $\delta$ breaks the sublattice symmetry and is introduced to lift the four band-touching points at the corners of the Brillouin zone. The Hamiltonian Eq.~\eqref{eq.H1} results in a flat and dispersive band, as shown in Fig.\,\ref{fig:BS_CB}(b) for reasonably generic choices of parameters.

\subsubsection{Model B}
\begin{figure}[htb]
    \centering
    \includegraphics[width=0.91\columnwidth]{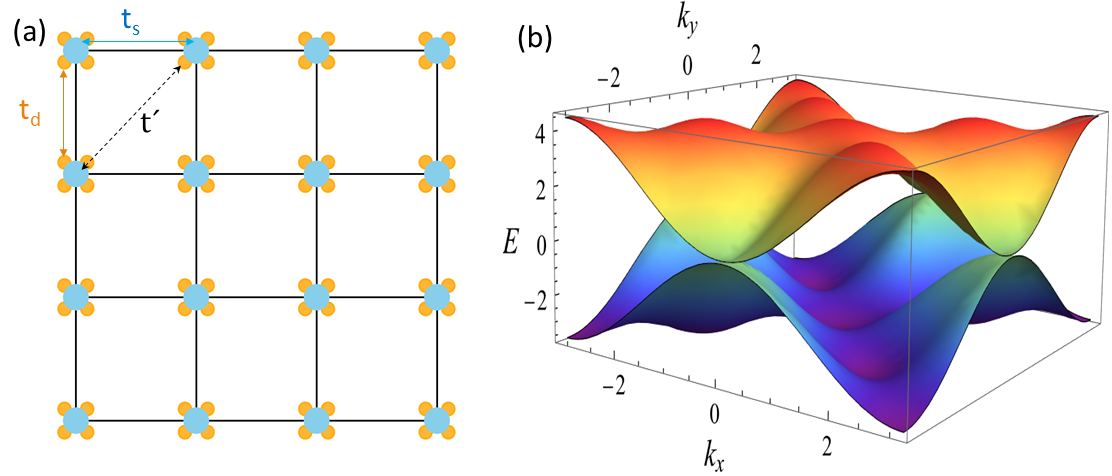}
    \caption{Model B described by the Hamiltonian Eq.~(\ref{eq.H2}). (a) Schematic diagram of the tight-binding square lattice containing $s$ (blue) and $d_{xy}$ (yellow) orbitals on each site and hopping parameters $t_s$, $t_d$, and $t^\prime$ denoting the intra- and inter-orbitals hopping amplitudes, respectively. (b) Energy dispersion for $t_s=1$, $t_d=-1$, $t^\prime=1$, $\mu_s=-0.5$, and $\mu_d=-0.4$, same as in Ref.~\cite{Chen23}.}
    \label{fig:BS_HM}
\end{figure}

For Model B we choose a two-orbital model consisting of an $s$ and a $d_{xy}$ orbital on each site of a square lattice, such that the kinetic Hamiltonian in the basis $ (c_{s, \mathbf{k}}, c_{d, \mathbf{k}})^T$ takes the form \cite{Chen23,Huang25}
\begin{align}
  H^\text{B}_{\mathbf{k}} = 
   \begin{pmatrix}
   \epsilon_{s, \mathbf{k}} & \lambda_{\mathbf{k}} \\
   \lambda^*_{\mathbf{k}} &   \epsilon_{d, \mathbf{k}} 
      \end{pmatrix} ,
       \label{eq.H2}
\end{align}
where $\epsilon_{i,\mathbf{k}} = -2\, t_{i} (\cos{k_x} + \cos{k_y)} - \mu_i$ (with $i \in \{ s, d \}$) denotes the intraorbital dispersion and $\lambda_{\mathbf{k}} = 4\, t^\prime \sin {k_x} \sin{k_y}$ represents a momentum-dependent inter-orbital mixing, see Fig.\,\ref{fig:BS_HM}(a). This model generically results in two dispersive bands as shown in Fig.\,\ref{fig:BS_HM}(b). Here we choose parameters in a way that it gives a sizable interband velocity, see further Sec.~\ref{sec:theory_Meissner}. 
We also note that a negative superfluid weight has been reported in this model~\cite{Chen23, Huang25}. It is interesting to investigate if this negative superfluid weight results from odd-frequency pairing.
We note that both systems described by the Hamiltonians in Eqs.~(\ref{eq.H1}, \ref{eq.H2}) preserve time-reversal symmetry in the normal state.
\subsection{Superconducting pairing correlations}
\label{sec:theory_SC}
To study superconductivity in general two-band models, including in Models A and B discussed above, we work in the band basis, where the kinetic energy in the normal state is diagonal. In the band basis, with the Nambu spinor $ \psi_{\mathbf{k}}= (c_{1,\mathbf{k},\uparrow}, c_{2,\mathbf{k},\uparrow}, c^\dagger_{1,-\mathbf{k},\downarrow}, c^\dagger_{2,-\mathbf{k},\downarrow})^T$, where $i=1,2$ denotes the two bands, the Bogoliubov-de Gennes (BdG) Hamiltonian is given by
\begin{align}
  \mathcal{H}^\text{BdG}_{\mathbf{k}} = 
   \begin{pmatrix}
     H_{N, b} (\mathbf{k}) & \hat{\Delta}_b \\ \\
     \hat{\Delta}_b^\dagger & -H^T_{N, b} (-\mathbf{k})
      \end{pmatrix}.
      \label{eq:BdG}
\end{align}
Here $H_{N, b} (\mathbf{k}) =\begin{pmatrix}
    E_1 & 0 \\
    0 & E_2
\end{pmatrix}$ is a diagonal matrix containing the band, or eigenenergies, $E_{\mathbf{k}i}$ (for Models A and B obtained after diagonalizing the matrices in Eq.\,(\ref{eq.H1}) and Eq.\,(\ref{eq.H2}), respectively), while $U_{\mathbf{k}}$ is the unitary matrix containing the associated eigenvectors $\psi_{\mathbf{k}i}$. The pairing matrix, originally assumed to be $\hat{\Delta}^\text{orb}$ in orbital space, can be transformed to the band basis through the transformation $ \hat{\Delta}_b=  U^{-1}_{\mathbf{k}}\, \hat{\Delta}^\text{orb}\,U^{*}_{-\mathbf{k}}$. 
Even if we just assume simple intraorbital pairing, possibly even only in one orbital, we generically generate both intra- and interband pairing following this transformation, i.e., $\hat{\Delta}_b = \begin{pmatrix}
    \Delta_{1, \mathbf{k}} & \Delta_{12, \mathbf{k}} \\
    \Delta_{21, \mathbf{k}} & \Delta_{2, \mathbf{k}} 
\end{pmatrix}$.
Here, interband pairing, $\Delta_{12, \mathbf{k}}$, is directly proportional to the interorbital hybridization and a finite difference in the intraorbital pairing amplitudes \cite{BB13,TCB20}. Thus, interband pairing is ubiquitous in multiband systems and is also present in both Models A and B, as soon as the pairing within each orbital is slightly different.
For the rest of this work, we focus only on conventional spin-singlet $s$-wave (with no $k$-dependence) superconductivity, as that is most common in nature. 

To obtain the superconducting pair correlator, we extract the Gorkov Green's function of the BdG Hamiltonian in Eq.\,(\ref{eq:BdG}) as
\begin{align}
    \mathcal{G} = (i \omega_n -\mathcal{H}^{\text{BdG}}_\mathbf{k})^{-1} \equiv \begin{pmatrix}
        G  & F \\
        \bar{F} & \bar{G}
    \end{pmatrix},
\end{align}
where $\omega_n = (2n +1)/ k_B T$ is the fermionic Matsubara frequency, with $T$ temperature, and the diagonal elements are the normal part of the Green's function, while the off-diagonal elements correspond to the anomalous Green's function. The bar corresponds to the Green's function in the hole sector. 

The anomalous part of the Green's function captures the superconducting pair correlations and can be further decomposed into even- and odd-frequency parts: $F= F^e + F^o$. 
The odd frequency part of the anomalous Green's function reads as
\begin{align}
   F^o =  \frac{1}{D} \begin{pmatrix}
       0 & i \omega \, (E_1 - E_2)\, \Delta \\
       -i \omega \, (E_1 - E_2)\, \Delta & 0
    \end{pmatrix},
    \label{eq:Fodd}
\end{align}
where we have set $\Delta=\Delta_{12, \mathbf{k}} = \Delta_{21, \mathbf{k}}$ for the interband pairing amplitude and $D = ((i\omega)^2- \xi^2_{+})^2 \, ((i\omega)^2- \xi^2_{-})^2$\, with $\xi_{\pm}$ being the two positive eigenvalues of the BdG Hamiltonian in Eq.\,(\ref{eq:BdG}), with $\xi_{+}>\xi_{-}$. As we see from Eq.\,(\ref{eq:Fodd}), $F^o$ is odd in frequency ($\omega$) and, notably, only depends on the interband pairing amplitude $\Delta$, not on the intraband pairing terms $\Delta_{1,2}$. This is expected since intraband pairing per definition is even in frequency. Therefore, to obtain nonzero $F^o$ in a two-band spin-singlet $s$-wave bulk superconductor, only a finite interband pairing is required to be present (along with two non-degenerate bands). 
Due to the presence of interorbital hybridization in both Hamiltonians Eqs.~(\ref{eq.H1}) and (\ref{eq.H2}), they both naturally host odd-frequency pairing in the presence of finite and different intraorbital pairing in the two orbitals \cite{BB13,TCB20}.

For comparison we also extract the expression for the even-frequency part of the anomalous Green's function, which reads as
\begin{align}
   F^e =  \frac{1}{D}\begin{pmatrix}
     \lambda_1 & \alpha  \\
     \alpha &   \lambda_2
    \end{pmatrix},
    \label{eq:Feven}
\end{align}
where $\alpha=-\Delta\,(E_1 E_2 + \Delta^2-\Delta_1 \Delta_2-\omega^2)$ and $\lambda_i=   \Delta^2\Delta_{j}-\Delta_i(E_{j}^2 + \Delta_{j}^2 -\omega^2)$, where we have set $\Delta_{1,2}$ as momentum-independent $s$-wave intraband pairing amplitudes, i.e., $\Delta_{1,\mathbf{k}}= \Delta_1$ and $\Delta_{2,\mathbf{k}}= \Delta_2$. Unlike $F^o$, the even-frequency part of anomalous Green's function $F^e$ depends on both the intra- and interband pairing amplitudes.

\subsection{Meissner effect}
\label{sec:theory_Meissner}
The Meissner effect, one of the defining features of superconductivity, determines the response of a superconductor to the application of an external magnetic field. Within Kubo linear response theory, the  response current $J_\mu$ is related to the vector potential $\mathbf{A}$ of an applied magnetic field through,
\begin{align}
    J_\mu (\mathbf{q}, \omega_{e}) = -K_{\mu \nu} (\mathbf{q}, \omega_{e}) A_{\nu} (\mathbf{q}, \omega_{e}),
\end{align}
where $\mathbf{q}$ and $\omega_e$ denote the wave vector and angular frequency of the external vector potential. Here, $K_{\mu \nu}$ is the current-current correlation function, also known as the Meissner kernel, while $\mu$, $\nu$ represent the direction of current response and external vector potential, respectively. In general, $K_{\mu \nu}$ can take either sign. When $K > 0$, the magnetic field is suppressed, giving the traditional diamagnetic Meissner response, while $K < 0$ corresponds to a paramagnetic Meissner response, which attracts magnetic field, thus ultimately destroying superconductivity. Within linear response theory, the response function $K_{\mu \nu}$ can be calculated as~\cite{Mahan2000,MO15,LLT17}
\begin{align}
    K_{\mu \nu} (\mathbf{q}, \omega_{e}) = \braket{J^P_{\mu} (\mathbf{q}, \omega_{e})\, J^P_{\mu} (-\mathbf{q}, \omega_{e})} + \braket{J^D_{\mu \nu} (\mathbf{q}, \omega_{e})},
\end{align}
where the expectation values are taken with respect to the unperturbed Hamiltonian. Here, $J^{P}$ and $J^{D}$ are the paramagnetic and diamagnetic current operators, respectively. They are obtained by including the vector potential in the Hamiltonian through the Peierl's substitution $\mathbf{k} \rightarrow \mathbf{k}-\mathbf{A}$ and expanding up to second order in $\mathbf{A}$ and then differentiating with respect to $A_{\nu}$. The constant and linear terms in $\mathbf{A}$ in the resulting expression are then identified as paramagnetic and diamagnetic current, respectively. 

The Meissner effect is the response to a static, uniform magnetic field, expressed through the limit $\lim_{\mathbf{q}\rightarrow0}\,\lim_{\omega_e \rightarrow 0} K_{\mu \nu} (\mathbf{q},\omega_{e})$, where the order of the limits is important \cite{SZ93,CB22}. The Meissner response, or kernel, can then be expressed in terms of Green's functions and first quantized current operators as
{\small{\begin{align}
    K_{\mu \nu} (\mathbf{q} =0, \omega_{e}=0)  = T \sum_{\mathbf{k}, i \omega} \Tr[ \hat{G}  \hat{J}^P_{\mu} \hat{G}  \hat{J}^P_{\nu} + \hat{F}  \hat{\bar{J}}^P_{\mu} \hat{\bar{F}}  \hat{J}^P_{\nu} + \hat{G}\hat{J}^D_{\nu}],
\end{align}}}
where $\hat{J}^P$ and $\hat{\bar{J}}^P$ are the current in the particle and hole sector, respectively.
Since we are interested in investigating the odd- and even-frequency contributions to the Meissner kernel, we only need to consider those terms that involve the anomalous Green's functions,
\begin{align}
    K^{S}_{\mu\nu} = T \sum_{\mathbf{k}, i \omega} \Tr[\hat{F}  \hat{\bar{J}}^P_{\mu} \hat{\bar{F}}  \hat{J}^P_{\nu} ]
    \label{eq:Ks},
\end{align}
which we can further divide into the even- and odd-frequency contributions
\begin{align}
    K^{S}_{\mu\nu} &= \mathcal{K}^{e}_{\mu\nu} + \mathcal{K}^{o}_{\mu\nu}  \nonumber \\ 
    &= T \sum_{\mathbf{k}, i \omega} \Tr[\hat{F}^e\hat{\bar{J}}^P_{\mu} \hat{\bar{F}}^e  \hat{J}^P_{\nu} ] +  T \sum_{\mathbf{k}, i \omega} \Tr[\hat{F}^o\hat{\bar{J}}^P_{\mu} \hat{\bar{F}}^o\hat{J}^P_{\nu} ], \label{eq:Ks}
\end{align}
For the last equality, we have used the fact that all mixed terms, $\hat{F}^e\hat{\bar{J}}^P_{\mu} \hat{\bar{F}}^o  \hat{J}^P_{\nu}$ and $\hat{F}^o\hat{\bar{J}}^P_{\mu} \hat{\bar{F}}^e \hat{J}^P_{\nu}$, are traceless and thus do not contribute.

To proceed we note that the current operator is related to the velocity matrix by $J^{ij} _\mu = e \,  V^{ij} _\mu$, where $i,j$ denotes the different bands in the normal state. The elements of velocity matrix in the band basis read as \cite{Rossi21, Torma23}
\begin{align}
    V^{ij} _\mu= \delta_{ij}\, \partial_{\mu} E_{\mathbf{k}i} + (E_{\mathbf{k}i} -E_{\mathbf{k}j}) \braket{\partial_{\mu} \psi_{\mathbf{k}i} |  \psi_{\mathbf{k}j}}, \label{eq:interbandvelocity}
\end{align}
where $\partial_{\mu} \equiv \partial_{\mathbf{k}_\mu}$ and $E_{\mathbf{k}i}$ and $\psi_{\mathbf{k}i}$ denote the eigenenergy and eigenvector, respectively, of the $i$-th Bloch band.  The diagonal velocity elements correspond to the usual group velocity, while the off-diagonal elements denote the interband velocity that signifies coherent transport of electrons between different bands. In particular, the object $\braket{\partial_{\mu} \psi_{\mathbf{k}i} |  \psi_{\mathbf{k}j}}$ in Eq.\,(\ref{eq:interbandvelocity}) is the non-Abelian Berry connection of two Bloch bands, which encodes the quantum geometric character of the Bloch states. As a consequence, the Meissner response may not only have the usual contributions from the dispersion, but in multiband systems also from the quantum geometric character of the Bloch states.
Further, in the presence of time-reversal symmetry, the velocity matrices in the particle and hole sector are related by $V^{ij}_{\mu} = \bar{V}^{ij}_{\mu}$ for all $i,j$. Therefore, the current operators in the band basis are related by $J^P = \bar{J}^P $.

We can simplify Eq.\,(\ref{eq:Ks}) by performing the matrix multiplications and taking the trace. This results in the Meissner kernels $\mathcal{K}^{e/ o}$ taking the generic form
\begin{align}
    \mathcal{K}^{e/ o} = \frac{ a^{e/ o} + b^{e/ o}\, (i \omega)^2 + c^{e/ o} \, (i \omega)^4 }{D}, 
    \label{eq:MKdef}
\end{align}
where $a^{e/ o}, b^{e/ o},$ and $c^{e/ o}$ are the coefficients for the even- or odd-frequency contributions, respectively. These coefficients are frequency-independent, but still functions of the wave vector $ \mathbf{k} $ and the system's parameters. It is the product of two anomalous Green's function in Eq.\,(\ref{eq:Ks}) that ensures that the resulting expressions contain only even powers of $ i \omega$.  
For an analytical treatment, it is useful to further simplify Eq.~\eqref{eq:MKdef}. This can be done by decomposing the Meissner kernel into its intra- and inter-band processes as~\cite{MO15,Schmidt20,PB21}
\begin{align}
    \mathcal{K} &= \mathcal{K}^{\text{intra}} + \mathcal{K}^{\text{inter}} \\
    \mathcal{K}^{\text{intra}} & = \sum_{\mathbf{k}, i\omega}\frac{\mathcal{K}_+}{((i \omega)^2- \xi^2_{+})^2} + \frac{\mathcal{K}_-}{((i \omega)^2- \xi^2_{-})^2}, \label{eq:M_freq_sum1} 
    \\
    \mathcal{K}^{\text{inter}} & = \sum_{\mathbf{k}, i\omega}\frac{\mathcal{K}_{\pm}}{((i \omega)^2- \xi^2_{+}) \,((i \omega)^2- \xi^2_{-})} \label{eq:M_freq_sum2} .
\end{align}
Here $\mathcal{K}^{\text{intra}}$ includes only 
terms with poles at one of the band energies, or BdG eigenvalues, $\xi_+$ or $\xi_-$, while in $\mathcal{K}^{\text{inter}}$ both of these band energies appear multiplied, hence it necessarily encodes interband processes. However, note that both $\mathcal{K}^{\text{intra}}$ and $\mathcal{K}^{\text{inter}}$ contain contributions from both the diagonal and off-diagonal elements of the velocity matrix in Eq.(\ref{eq:interbandvelocity}). Thus, both $\mathcal{K}^{\text{intra}}$ and $\mathcal{K}^{\text{inter}}$ incorporate contributions from both conventional (intraband velocity) and geometric terms (interband velocity). 

The coefficients $\mathcal{K}_{+}, \mathcal{K}_{-}$, and $\mathcal{K}_{\pm}$ in Eqs.~(\ref{eq:M_freq_sum1}-\ref{eq:M_freq_sum2}) can be expressed in terms of the coefficients $a, b,$ and $c$ in Eq.\,(\ref{eq:MKdef}) and the eigenvalues of the BdG Hamiltonian in Eq.~(\ref{eq:BdG}) as
\begin{align}
 \mathcal{K}_+ & = \frac{a + b \,\xi_{+}^2 + c \, \xi_{+}^4}{(\xi_{+}^2 - \xi_{-}^2)^2}, \label{eq:M_freq1}\\
    \mathcal{K}_- & = \frac{a + b \, \xi_{-}^2 + c \, \xi_{-}^4}{(\xi_{+}^2 - \xi_{-}^2)^2},\label{eq:M_freq2} \\ 
    \mathcal{K}_{\pm} & = -\frac{2 a + b \,(\xi_{+}^2 + \xi_{-}^2)+ 2 c \, (\xi_{+}^2 \xi_{-}^2)}{(\xi_{+}^2 - \xi_{-}^2)^2}, \label{eq:M_freq3}
\end{align}
where $a = a^{e/o}$ and so on.
The Matsubara summation in Eqs.\,(\ref{eq:M_freq_sum1}-\ref{eq:M_freq_sum2}) can further be performed using the identities~\cite{PB21}
\begin{align}
   &  T \sum_{i\omega} \frac{1}{((i \omega)^2- \xi_{\pm}^2)^2}  = \frac{C (\xi_{\pm}) + \eta^\prime (\xi_{\pm})}{2 \xi_{\pm}^2}, \\
   &  T \sum_{i\omega} \frac{1}{((i \omega)^2- \xi_{+}^2) ((i \omega)^2- \xi_{-}^2)}  = -\frac{C (\xi_{+}) - C (\xi_{-})}{(\xi_{+}^2 - \xi_{-}^2)}, \label{eq: Matsubarasum}
\end{align}
with the Fermi-Dirac distribution $\eta(\xi) = \frac{1}{2} \left( 1 - \tanh \frac{\beta \xi}{2} \right)$, its derivative  $\eta^\prime(\xi)=-\frac{\beta}{4} \sech\frac{\beta \xi}{2}$, and $C (\xi) = \frac{1}{2 \xi} \tanh \frac{\beta \xi}{2}$. 
Finally, combining Eqs.\,(\ref{eq:M_freq1}-\ref{eq: Matsubarasum}), $\mathcal{K}^\text{intra}$ and $\mathcal{K}^\text{inter}$ are obtained individually for even- and odd-frequency contributions after summing over momentum $\mathbf{k}$ as

\begin{align}
    \mathcal{K}^\text{intra}&= \sum_{\mathbf{k}} \frac{a + b \,\xi_{+}^2 + c\, \xi_{+}^4}{2 ( \xi_{+}^2 -  \xi_{-}^2)^2 \,\xi_{+}^2} \, ( C (\xi_{+}) + \eta^\prime (\xi_{+})) \nonumber \\ 
     & + \quad  \frac{a + b \,\xi_{-}^2 + c \, \xi_{-}^4}{2 ( \xi_{+}^2 -  \xi_{-}^2)^2 \,\xi_{-}^2} \, ( C (\xi_{-}) + \eta^\prime (\xi_{-})) \label{eq:Kintra}\\
     \mathcal{K}^\text{inter}&=  \sum_{\mathbf{k}} \frac{2 a + b\, (\xi_{+}^2 + \xi_{-}^2) + 2 c \,\xi_{+}^2 \,\xi_{-}^2 }{( \xi_{+}^2 -  \xi_{-}^2)^3} \,(C (\xi_{+}) - C (\xi_{-}) ).\label{eq:Kinter}
 \end{align}
The signs of the coefficients $a, b,$ and $c$ are therefore crucial in determining the overall sign of the Meissner kernel, as we will highlight in our results in Sec.~\ref{sec:results}. 

\section{Results} 
\label{sec:results}
Having established the necessary formalism to determine the role of quantum geometry for the odd- and even-frequency pairing contributions to the Meissner response, we proceed to discuss our results for two band systems. We start with a general analytical analysis and continue by reporting numerical results for the two model systems introduced in Sec.~\ref{sec:theory_model} to obtain more detailed results.

\subsection{General analysis}
\label{results: GA}
To proceed analytically we first consider the simplifying case of pure interband pairing, i.e., when both $\Delta_{1,2}=0$, but $\Delta_{12} = \Delta_{21}=\Delta \neq 0$. 
After performing the matrix multiplications and taking the trace in Eq.\,(\ref{eq:Ks}), the expression for $ \mathcal{K}_{\mu \nu}^{e/o}$ then simplifies to
\begin{align}
   \mathcal{K}_{\mu \nu}^e & = T \sum_{k, i\omega} \frac{|\Delta |^2\, (|\Delta |^2 + E_1\,E_2 - (i \omega)^2)^2 \, (J^\text{conv} + J^\text{geom}) }{D}
     \label{eq:ke_intra}
\end{align}
and
\begin{align}
    \mathcal{K}^o_{\mu \nu} =  T \sum_{\mathbf{k}, i \omega} \frac{|\Delta|^2 \,(E_1-E_2)^2  (J^\text{conv}-J^\text{geom})}{D} (i \omega)^2, \label{eq:Ko_expression}
\end{align}
 where $J^\text{conv}= ( J^\mu_{11}\,J^\nu_{22} + J^\nu_{11}\,J^\mu_{22} )$ and $J^\text{geom}= ( J^\mu_{12}\,J^\nu_{21} + J^\mu_{21}\,J^\nu_{12}\, )$, are the contributions from the diagonal (conventional) current and off-diagonal (interband) currents or velocities, respectively. We here note that, while the interband current by itself is not gauge-invariant,  the products that appear in the response functions, i.e., $J^\mu_{12}\,J^\nu_{12}$ are gauge-invariant. Further, $J^\text{geom}$ simplifies to
\begin{align}
    J^\text{geom} = 2 (E_1 - E_2)^2\, \text{Re}[\langle{\partial_\mu \psi_{\mathbf{k}, i}| \psi_{\mathbf{k}, j} \rangle } \langle{\psi_{\mathbf{k}, j}| \partial_\nu \psi_{\mathbf{k}, i}\rangle }]\label{eq: Jgeom},
\end{align}
with $i \neq j$.
For a two-band system, $\ket{\psi_{\mathbf{k}, j}}\bra{\psi_{\mathbf{k}, j}} = 1-\ket{\psi_{\mathbf{k}, i}}\bra{\psi_{\mathbf{k}, i}} $ and thus
\begin{align}
   \langle{\partial_\mu \psi_{\mathbf{k}, i}| \psi_{\mathbf{k}, j} \rangle } \langle{ \psi_{\mathbf{k}, j}| \partial_\nu \psi_{\mathbf{k}, i}\rangle } & =  \langle{\partial_\mu \psi_{\mathbf{k}, i}}|\partial_\nu \psi_{\mathbf{k}, i} \rangle \nonumber \\
   &-\langle{\partial_\mu \psi_{\mathbf{k}, i}}| \psi_{\mathbf{k}, i}\rangle \langle{\psi_{\mathbf{k}, i}} |\partial_\nu \psi_{\mathbf{k}, i} \rangle \nonumber \\
   &= Q^i_{\mu \nu},
\end{align}
where $Q^i_{\mu \nu}$ is the quantum geometric tensor (QGT) of the Bloch band $i$. Therefore, $ J^\text{geom}$ in Eq.~(\ref{eq: Jgeom}) is related to the real part of the QGT, i.e., the quantum metric of the Bloch band $i$ (or equivalently Bloch band $j$ for a two-band system). 
This result holds exactly for a two-band system. For systems possessing more than two bands, $J^\text{geom}$ in Eq.~(\ref{eq: Jgeom}) cannot be expressed as the quantum metric of an individual band. Nevertheless, the quantum geometric character of the Bloch bands is still evident in $J^\text{geom}$ through the non-Abelian Berry connection between different Bloch bands in Eq.~(\ref{eq: Jgeom}).

We can proceed further by identifying from the expression for $\mathcal{K}_{\mu \nu}^e$ in Eq.~(\ref{eq:ke_intra}), the coefficients in Eq.~\eqref{eq:MKdef} as
\begin{align}
    a^{e} & = |\Delta |^2\, (|\Delta |^2 + E_1\,E_2)^2 \, (J^\text{conv} + J^\text{geom}), \\
    b^{e} & =  -2|\Delta |^2\, (|\Delta |^2 + E_1\,E_2) \, (J^\text{conv} + J^\text{geom}), \\
    c^{e} & = |\Delta |^2\, (J^\text{conv} + J^\text{geom}).
    \label{eq: even-coeffs}
\end{align}
For a general pairing state, i.e., when both $\Delta_{1,2} \neq 0$ and $\Delta \neq 0$, we obtain the results for $\mathcal{K}_{\mu \nu}^e$ similarly, but the expressions are lengthy and provide little insight and therefore we do not provide them here. Still, we conclude that, since all coefficients $a^e, b^e,$ and $c^e$ are nonzero, the overall signs of $\mathcal{K}^\text{inter/intra}$ for even-frequency pairing are determined by their interplay and will vary for different models and parameters.

The same identification of coefficients for the odd-frequency contribution to the Meissner kernel $\mathcal{K}_{\mu \nu}^o$ in Eq.~(\ref{eq:Ko_expression}) with Eq.~\eqref{eq:MKdef} results in
\begin{align} 
b^o &= |\Delta|^2 \,(E_1-E_2)^2  (J^\text{conv}-J^\text{geom}),\\
a^o&=c^o=0.
\label{eq:odd_coeff}
\end{align}
We additionally find that the numerator of the expression for $\mathcal{K}_{\mu \nu}^o$ in Eq.~(\ref{eq:Ko_expression}) remains the same for the general pairing case as for the pure interband case, i.e., it does not depend on the amplitudes of the intraband pairing.  Therefore, the expressions for $a^o, b^o$ and $c^o$ in Eq.~\eqref{eq:odd_coeff} are valid for all pairing scenarios. In particular, since only $b^o \neq 0$, we can in the following make definitive and general statements about how odd-frequency pairing influences the Meissner response, including its conventional and geometric contributions. For the latter purpose it is useful to write $b^o = (b^o)^\text{conv} + (b^o)^\text{geom}$ with  
\begin{align}
    (b^o)^\text{conv} & = |\Delta|^2 \,(E_1-E_2)^2  \,J^\text{conv} \\
    (b^o)^\text{geom} &= -|\Delta|^2 \,(E_1-E_2)^2  \,J^\text{geom} <0.
\end{align}
Note that $(b^o)^\text{geom}$ is always negative, at least for two-band systems, since $J^\text{geom}$ is directly proportional to the quantum metric and thus, positive.

We start with the interband contribution $\mathcal{K}^\text{inter}$ to the Meissner kernel expressed in Eq.~(\ref{eq:Kinter}) for odd-frequency pairing. With $a^o=c^o=0$ and noting that $(C (\xi_{+}) - C (\xi_{-}) )$ is negative, it is the sign of $b^o$ that determines the overall sign of $\mathcal{K}^\text{inter}$ for odd-frequency pairing. In particular, if $b^{o}<0$, then $\mathcal{K}^\text{inter}>0$ and thus we have a diamagnetic interband Meissner response for the odd-frequency channel. Therefore, the geometric contribution to $\mathcal{K}^\text{inter}$ for odd-frequency pairing is \textit{always diamagnetic}. 
On the other hand, $(b^o)^\text{conv}$ can both be positive or negative depending on the band structure. If the two bands have the same curvature, meaning $J^\mu_{11}\,J^\nu_{22} >0$, we have $J^\text{conv}>0$ and thus $(b^o)^\text{conv}>0$, leading to the interband Meissner response being paramagnetic. In contrast, for two inverted bands, for which $J^\mu_{11}\,J^\nu_{22} <0$, we have $(b^o)^\text{conv}<0$, which leads to a diamagnetic interband response. This behavior of the conventional contribution of the interband Meissner kernel is consistent with the results of Ref.~\cite{PB21}, but where the geometric contribution was not included. 

Moving on to the intraband contribution $\mathcal{K}^\text{intra}$ to the Meissner kernel expressed in Eq.~(\ref{eq:Kintra}), we again find that the sign of $b^o$ alone determines the overall sign of the odd-frequency contribution to the Meissner response. This is because $C (\xi_{\pm})$ in Eq.~(\ref{eq:Kintra}) is always positive, while $\eta^\prime(\xi_{\pm})$ is always negative but negligible for a gapped superconductor \cite{PB21}, as is the case considered here. Thus, since $(b^o)^\text{geom}$ is always negative, the geometric contribution to $\mathcal{K}^\text{intra}$ is \textit{always paramagnetic} for odd-frequency pairing. On the other hand, for the conventional part, since $(b^o)^\text{conv}$ can both be positive or negative, we find that when the two bands have the same curvature, the intraband Meissner response is diamagnetic, while it is paramagnetic for two inverted bands, again consistent with Ref.~\cite{PB21}.
To conclude, we find that for odd-frequency pairing, the geometric contribution has definitive behavior. Its interband part $\mathcal{K}^\text{inter}$ is always diamagnetic, while its intraband part $\mathcal{K}^\text{intra}$ is always paramagnetic. Analytically we cannot determine the relative strength of the $\mathcal{K}^\text{inter}$ and $\mathcal{K}^\text{intra}$ contributions, but with odd-frequency
pairing itself generated by interband pairing, we may anticipate that interband processes are easily dominating in many odd-frequency systems. We confirm this numerically in the next subsection.
Thus a superconductor dominated by the geometric response, as in a flat band system, can host substantial, or even purely, odd-frequency pairing and still have a diamagnetic Meissner effect, as long as its interband Meissner kernel is dominating. 
This should allow for the viable existence of bulk odd-frequency superconductivity in flat bands systems, as we also demonstrate numerically in the next subsection. This conclusively answers the question about the geometric contribution to the Meissner effect of odd-frequency pairing. 
The conventional Meissner response has a less simple behavior and instead depends on the curvature of the bands. For even-frequency pairing, we find no simple analytical predictions. 

\subsection{Numerical results for Model A}
\label{results:Model A}
\begin{figure}[t!]
    \centering
    \raisebox{1ex}{ \includegraphics[width=\columnwidth]{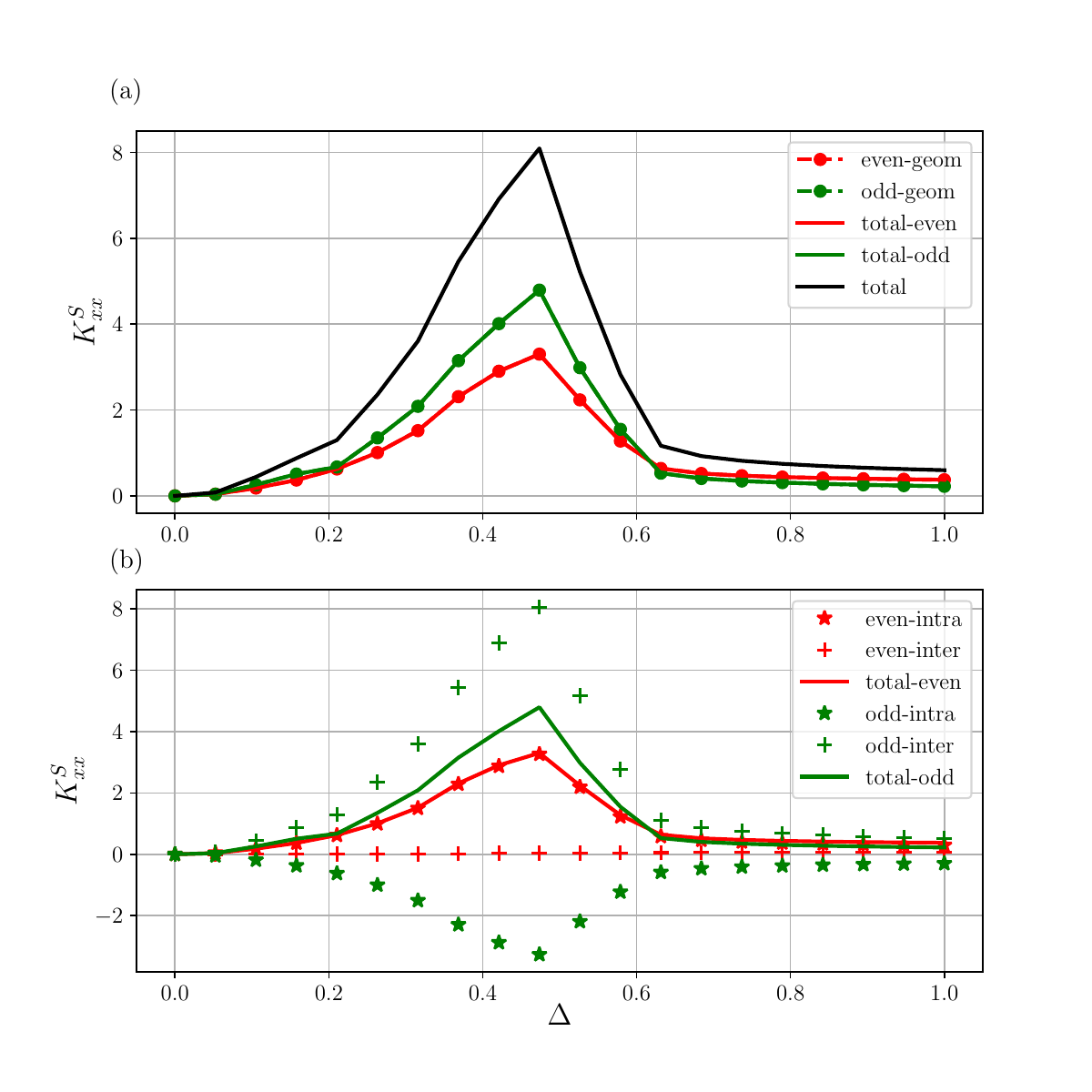}}
    \caption{ (a) Even- and odd-frequency pairing contributions to the Meissner response from Eq.~(\ref{eq:Ks}) and their corresponding geometric contributions as a function of interband pairing amplitude $\Delta$ for pure interband pairing in Model A given by the Hamiltonian in Eq.~(\ref{eq.H1}).  (b) Same as in (a) but instead divided into intra- and interband contributions.}
    \label{fig:pureinterband_CB}
\end{figure}
We complement our analytical results in the previous subsection with extensive numerical modeling. Since we consider isotropic pairing states, we can without loss of generality choose the $x$-component of the Meissner response, i.e., $K_{xx}$ to discuss our results. First, we focus on Model A, which hosts one flat band and one dispersive band, described by the Hamiltonian in Eq.(\ref{eq.H1}), where we put the flat band at the Fermi level to maximize its contribution to the Meissner response.
We start by considering the simplest scenario for generating odd-frequency pairing, i.e.~assuming that the intraband pairing amplitudes are zero and only keeping the interband pairing amplitude $\Delta$ nonzero. 
The even- and odd-frequency contributions to the Meissner response and the corresponding geometric contributions as a function of interband pairing amplitude $\Delta$ is shown in Fig.~\ref{fig:pureinterband_CB}(a). We find a large diamagnetic response for both the even- and odd-frequency contributions, entirely given by the geometric contribution. Since the band at the Fermi level is entirely flat with vanishing group velocity, the absence of a conventional contribution is expected. Already this result establishes that odd-frequency pairing can carry a large diamagnetic Meissner response in the flat band limit, coming entirely from the geometric character of the underlying normal state.

To also better understand our analytical results in Sec.~\ref{results: GA}, we further analyze the even- and odd-frequency Meissner response in terms of their intra- and interband contributions in Fig.~\ref{fig:pureinterband_CB}(b). From Fig.~\ref{fig:pureinterband_CB}(a) we already know that all contributions are geometric in origin. Figure~\ref{fig:pureinterband_CB}(b) further shows that the odd-frequency interband contribution is diamagnetic, while the intraband contribution is paramagnetic. This is in full agreement with our analytical results. We further find that the diamagnetic interband contribution is heavily dominating, thus giving an overall diamagnetic Meissner effect from the odd-frequency pairs. When it comes to the even-frequency pairing, it is entirely carried by an diamagnetic intraband contribution. Hence, the odd- and even-frequency pairs both help to contribute to a stable superconducting state through a finite quantum metric.

\begin{figure}[t!]
    \centering
    \includegraphics[width=\columnwidth]{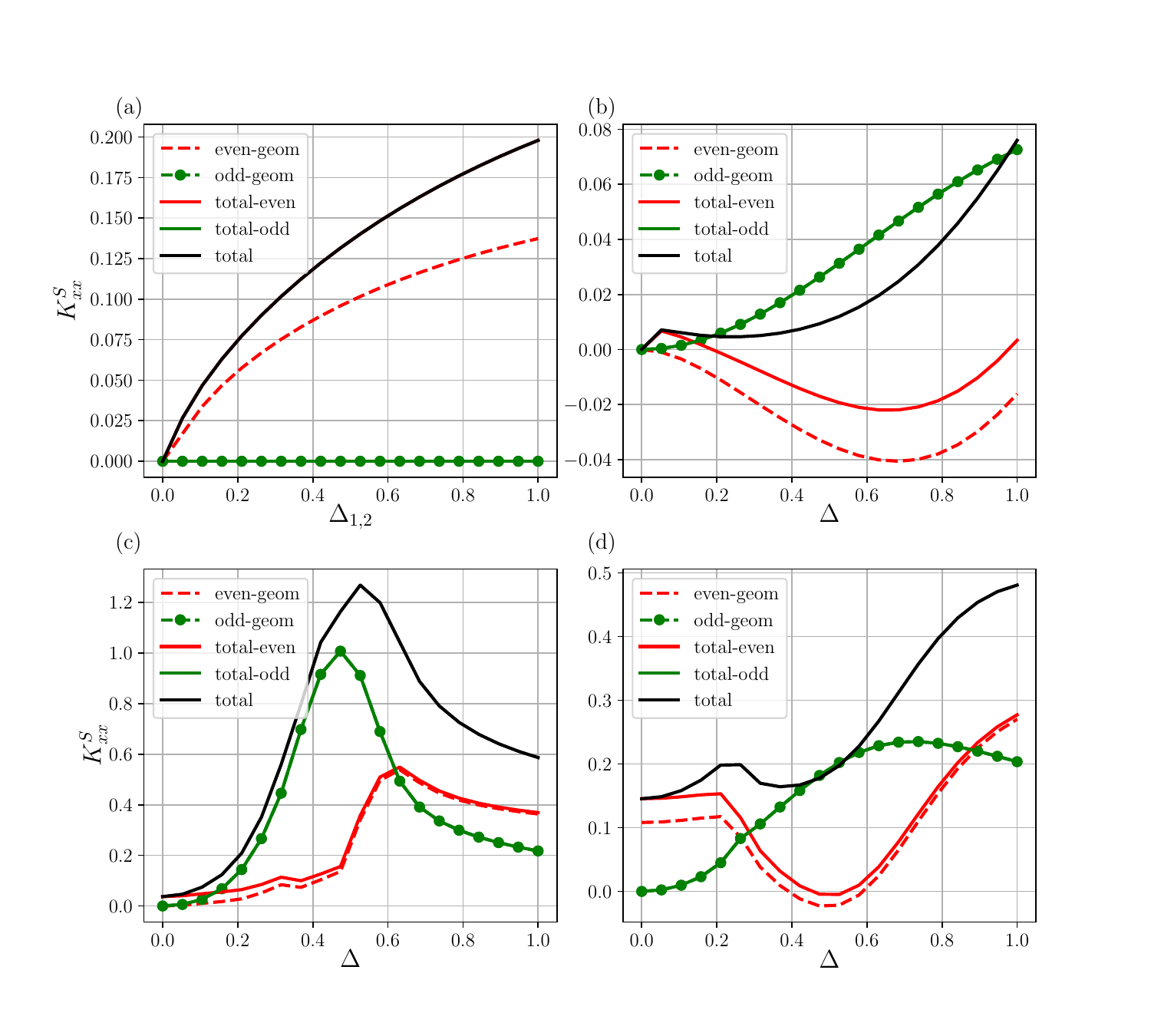}
    \caption{Even- and odd-frequency pairing contributions to the Meissner responses from Eq.~(\ref{eq:Ks}) and their geometric contributions in Model A given by the Hamiltonian in Eq.~(\ref{eq.H1}). In (a) as a function of intraband pairing amplitudes $\Delta_1=\Delta_2$ for $\Delta=0$, while as a function of interband pairing strength $\Delta$ for (b) $\Delta_1=0.0$, $\Delta_2=0.5$, (c)  $\Delta_1=0.5$, $\Delta_2=0.0$, and (d)  $\Delta_1=0.5$, $\Delta_2=0.2$. }
    \label{fig:CB_GP_P2}
\end{figure}

Finally, we also consider finite intraband pairing amplitudes in Model A.
In Fig.~\ref{fig:CB_GP_P2}(a), we focus on the case of pure intraband pairing with $\Delta_1 = \Delta_2\neq 0$ and $\Delta =0$. In this case, there is no odd-frequency pairing present and the Meissner effect entirely originates from even-frequency pairs, with a dominating diamagnetic geometric contribution but also with a finite conventional contribution. The latter we attribute to the imposed pairing also in the dispersive band as both $\Delta_{1,2}$ are finite.
Fig.~\ref{fig:CB_GP_P2}(b,c) then display the different contributions to the Meissner response as a function of $\Delta$ for the cases where one of the intraband pairing amplitude are zero. In Fig.~\ref{fig:CB_GP_P2}(b) the pairing set to zero in the flat band (band 1) but finite in the dispersive band (band 2), which generates a relatively small Meissner effect, as can be expected. Most notably, the diamagnetic Meissner effect is entirely generated by the odd-frequency pairs and also coming from their geometric contribution, as the even-frequency contribution is paramagnetic. 
In Fig.~\ref{fig:CB_GP_P2}(c), where the intraband pairing is instead in the flat band, but not in the dispersive band, there is a diamagnetic response also from the even-frequency pairs, but it is still the odd-frequency geometric contribution that gives the dominant Meissner effect. 
Finally, in Fig.~\ref{fig:CB_GP_P2}(d) we investigate the most general case, where $\Delta_{1,2}\neq 0$ with $\Delta_1>\Delta_2$, since that is reasonable to assume for a flat band at the Fermi level. Also here we find that the odd-frequency contribution to the Meissner response is not only diamagnetic, but also the dominant contribution for most parameters and its stems entirely from its geometric part. 
To summarize, these results establish that odd-frequency pairing in flat-band superconductors generates a strong diamagnetic Meissner effect stemming from the finite quantum metric of the system and interband processes in the Meissner kernel. It is even possible to generate scenarios where the even-frequency pairing generates diminishing or even paramagnetic Meissner effect, such that a stable superconducting state is only achieved because of odd-frequency pairs. This clearly demonstrates the feasibility of dominating, or even pure, odd-frequency pairing in flat band superconductors with a diamagnetic Meissner effect.
 \subsection{Numerical results for Model B}
 \label{results:Model B}
 \begin{figure}[t!]
    \centering
    \includegraphics[width=\columnwidth]{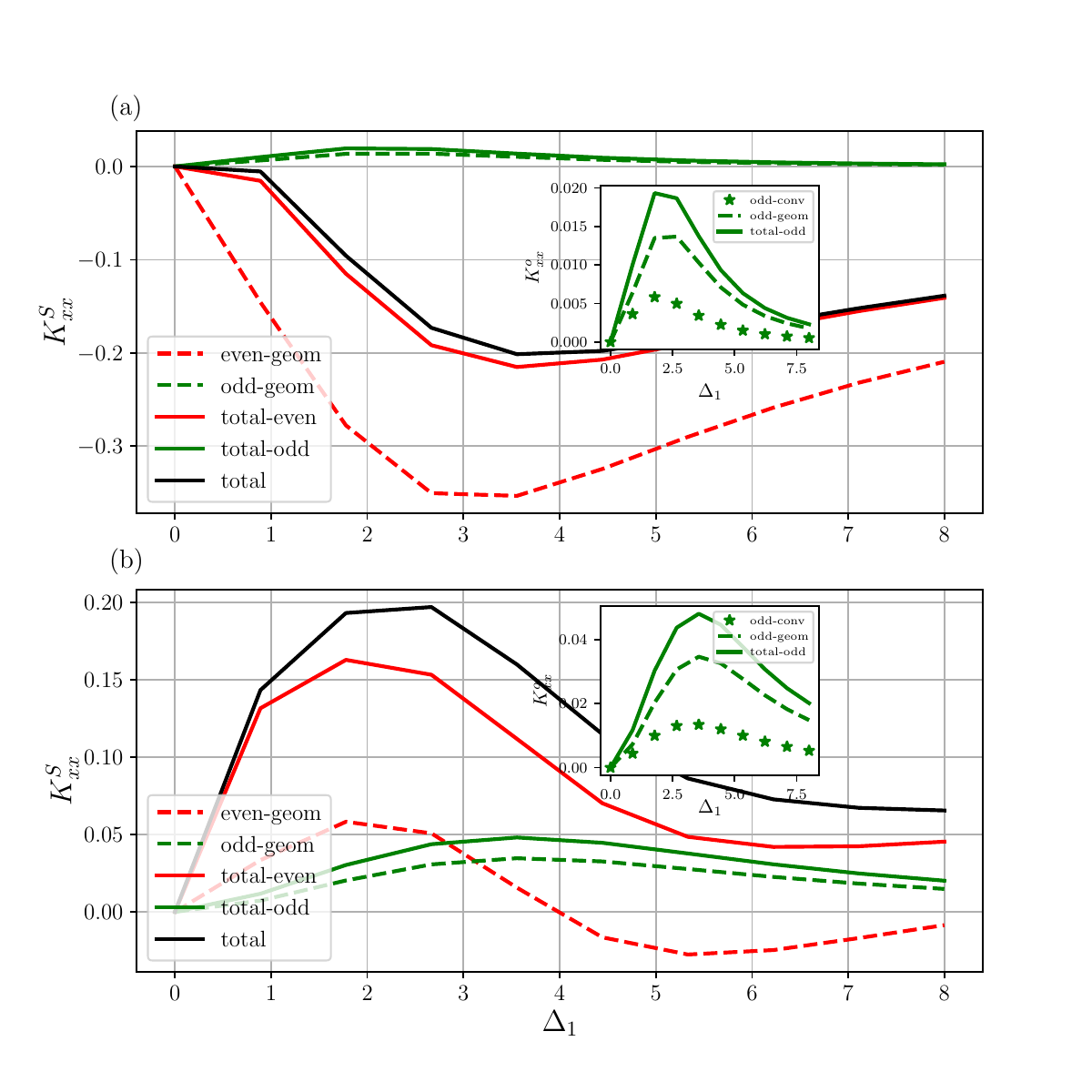}
    \caption{Even- and odd-frequency pairing contributions to the Meissner response from Eq.~(\ref{eq:Ks}) and their geometric contributions in Model B given by the Hamiltonian in Eq.~(\ref{eq.H2}) as  a function of intraband pairing strength for (a) intraband pairing $\Delta_1=-\Delta_2$ and (b) $\Delta_2=0.1\Delta_1$, with in both cases $\eta=1$ for interband pairing. Insets show zoomed-in versions of the odd-frequency contributions to the Meissner responses, also with a minor conventional contribution.}
    \label{fig:HM_GP_P2}
\end{figure}

To complement and contrast the results of Model A we also study Model B, described by the Hamiltonian in Eq.\,(\ref{eq.H2}) and hosting two dispersive bands. For the superconducting state, we consider two different scenarios, where the intraband pairing strengths are (a)  equal but with opposite phase: $\Delta_1 = -\Delta_2$ and (b) different but in the same phase: $ \Delta_1\neq \Delta_2$ with $\Delta_1 > \Delta_2$. For both cases we also assume an interband pairing that follows the $k$-dependence of the interorbital hybridization, i.e.~$\Delta = \eta \,\Delta_1\,\sin {k_x}\,\sin {k_y}$, where $\eta$ regulates the strength of the interband pairing. Beyond these being reasonable assumptions for superconductivity, we are motivated to study these particular pairing states by the fact that the geometric contribution to the superfluid weight has been shown to be negative for the (a) scenario~\cite{Chen23, Huang25}. This opens up for the intriguing additional question if that is due to even- or odd-frequency pairs?
We here note that Model B hosts odd-frequency pairing as soon as interband pairing is present \cite{BB13}. In addition, scenario (a) results from opposite intraorbital pairing amplitudes~\cite{Chen23}, which generates maximal intraband pairing asymmetry that further promotes odd-frequency pairing \cite{BB13,TCB20}.
 
The even- and odd-frequency contributions to the Meissner response and their corresponding geometric contributions for both scenarios (a) and (b) are plotted in Fig.~\ref{fig:HM_GP_P2}(a,b) respectively. In Fig.~\ref{fig:HM_GP_P2}(a) we find that the total Meissner response is paramagnetic, in agreement with Refs.~\cite{Chen23,Huang25}. Here we find that this occurs because of a dominant paramagnetic even-frequency geometric contribution, while the odd-frequency pairing gives a diamagnet response. Hence, it is not odd-frequency pairing that destabilizes the superconducting state as one could naively have expected, but the paramagnetic response is driven by even-frequency Cooper pairs.
In Fig.~\ref{fig:HM_GP_P2}(b) the total Meissner response turns diamagnetic because both even- and odd-frequency pairing contribute diamagnetically, although the even-frequency geometric contribution turns negative at strong coupling. Notably, the odd-frequency Meissner response is also dominated by the quantum geometry as seen in both insets, despite the bands being dispersive. We further note that the even-frequency response is dominating in both scenarios. The results in Fig.~\ref{fig:HM_GP_P2} demonstrate that odd-frequency pairing generates a diamagnetic Meissner response from the quantum metric, fully in line with the results from the flat band system in Figs.~\ref{fig:pureinterband_CB}-\ref{fig:CB_GP_P2}. Based on our analytical results we further conclude that this diamagnetic odd-frequency response is due to dominating interband processes.

\section{Conclusions} 
\label{sec:conclusions}
Odd-frequency superconducting pairing is ubiquitous in multiband systems, just requiring finite interband pairing \cite{BB13,PB21}.
However, odd-frequency superconductors have traditionally been assumed to host a paramagnetic Meissner response \cite{Hoshino14,AS15,PRX15,LSR17,KPG20}, and thus odd-frequency pairing has often been assumed to be detrimental to superconductivity.
At the same time, the quantum geometric structure of the normal-state Bloch bands, through the interband velocity, has recently been shown to significantly influence the superfluid weight or, equivalently, the Meissner response in multiband systems \cite{Rossi21,Torma23,Barlas24}. 
In this work, we answer the question how the quantum geometric structure affects how odd-frequency pairing contributes to the Meissner response. This question is particularly important for flat band superconductors, where quantum geometric effects dominate, while also easily generating large odd-frequency pairing correlations.

By dividing up the Meissner response into its contributions from the conventional current, i.e.~electron dispersion, and geometric contribution, shown to be generated by the quantum metric for a two-band superconductor, as well as it parts coming from intra- and interband processes, we are able to analytically establish definite behavior for the odd-frequency pairing: the geometric contribution in interband processes is always diamagnetic, while the geometric contribution in intraband processes is always paramagnetic. 
While we cannot analytically compare these two contributions, it is sensible to assume that interband processes may dominate for odd-frequency pairing, since interband processes are also the origin of odd-frequency pairs, thereby giving a diamagnetic odd-frequency Meissner response. 
For the conventional contribution, the odd-frequency response is instead dependent on the sign of the curvature of the bands, and hence disappears for flat bands. 
We also find that the even-frequency Meissner response does not follow any definite behavior, but is intricately dependent on the various strengths of the intra- and interband pairing.

We then proceed numerically by considering two simple, yet generic, two-band models to describe the normal state of prototype flat-band and dispersive systems, respectively, which both naturally host odd-frequency pairing.
For Model A, hosting a flat band and a dispersive band, we find that the odd-frequency contribution to the Meissner response is diamagnetic and a fully geometric contribution. This is achieved by the diamagnetic interband processes dominating the paramagnetic intraband processes in the Meissner response, fully aligning with our analytical findings. Importantly, the interband odd-frequency contribution can also easily be larger than the even-frequency contribution, even when there is substantial even-frequency intraband pairing, thus making odd-frequency interband pairs the primary driver for a diamagnetic Meissner effect. These results establish that the contributions of odd-frequency pairing to the Meissner response for a flat-band system is expected to be diamagnetic, fully generated by the finite quantum geometry, and possibly even the dominant contribution.

To complement the flat band results, we also study a Model B, hosting two dispersive bands with a finite quantum geometry. Here we also find that the odd-frequency pairing generates an diamagnetic Meissner response, primarily coming from the finite quantum metric. Interestingly, there exists parameter regimes for Model B where the Meissner effect is paramagnetic, but we find that that is instead due to even-frequency pairs producing a paramagnetic response, which dominates the smaller odd-frequency diamagnetic response.

In summary, our results establish that the geometric contribution to the Meissner response for odd-frequency pairing is diamagnetic and generated by strong interband processes. In flat band systems this odd-frequency driven diamagnetic Meissner response can even dominate all even-frequency contributions. This conclusively answers the questions raised in the introduction of this work. 
In addition, our results also reveal flat band systems as a promising avenue for finding a purely odd-frequency superconductor, still with a stabilizing diamagnetic Meissner response. 
Furthermore, finite-momentum superconductors may be another fertile ground for searching for odd-frequency superconductivity, since finite-momentum superconductors have both been shown to be related to odd-frequency superconducting correlations \cite{CB21,CB22} and manifest  quantum geometric effects \cite{KY22,KY23,Barlas23}. 

\section{Acknowledgment} 
\label{sec:acknowledgement}
We thank Debmalya Chakraborty, Paramita Dutta, Rodrigo Arouca and Quentin Marsal for useful discussions related to the work. We acknowledge financial support from the Knut and Alice Wallenberg Foundation through the project grant KAW 2019.0068 and the European Union through the European Research Council (ERC) under the European Union’s Horizon 2020 research and innovation programme (ERC-2022-CoG, Grant agreement No.~101087096). Views and opinions expressed are however those of the author(s) only and do not necessarily reflect those of the European Union or the European Research Council Executive Agency. Neither the European Union nor the granting authority can be held responsible for them.

%

\end {document}